\documentclass[aps,prd,twocolumn,groupedaddress,showpacs,nofootinbib,amssymb]{revtex4}

\usepackage[T1]{fontenc}
\usepackage[latin1]{inputenc}
\usepackage{graphicx}
\usepackage[english]{babel}
\usepackage{graphicx}
\usepackage{bm}
\usepackage{amsmath}
\usepackage{amssymb}
\usepackage{amsfonts}
\usepackage{epsfig}
\usepackage{colordvi}
\usepackage{color}

\begin{document}
\title{Cosmological inflation in $F(R,\mathcal{G})$ gravity}
\author{Mariafelicia De Laurentis$^{1,2}$\footnote{e-mail address: mfdelaurentis@tspu.edu.ru}, 
Mariacristina Paolella $^{3,4}$\footnote{e-mail address: paolella@na.infn.it}, Salvatore Capozziello$^{3,4,5}$\footnote{e-mail address: capozzie@na.infn.it}}
\affiliation{$^{1}$Tomsk State Pedagogical University, ul. Kievskaya, 60, 634061 Tomsk, Russia}
\affiliation{$^{2}$National Research Tomsk State University,
Lenin Avenue, 36, 634050 Tomsk, Russia }
\affiliation{$^{3,4}$Dipartimento di Fisica, Universit\' a
di Napoli {}``Federico II'', Compl. Univ. di
Monte S. Angelo, Edificio G, Via Cinthia, I-80126, Napoli, Italy\\
INFN Sezione  di Napoli, Compl. Univ. di
Monte S. Angelo, Edificio G, Via Cinthia, I-80126, Napoli, Italy.}
\affiliation{$^5$Gran Sasso Science Institute (INFN), Via F. Crispi 7, I-67100, L' Aquila, Italy.}

\date{\today}

\begin{abstract}
Cosmological inflation is discussed in the framework of $F(R,{\cal G})$ gravity where $F$ is a generic function of the curvature scalar $R$ and the Gauss-Bonnet topological invariant 
$\cal G$. The main feature that emerges in this analysis is the fact that this kind of theory can exhaust all the curvature budget related to curvature invariants without considering derivatives of $R,$ $R_{\mu\nu}$, $R^{\lambda}_{\sigma\mu\nu}$ etc. in the action.  Cosmological dynamics results driven by two effective masses (lenghts) related to the $R$ scalaron and the $\cal G$ scalaron working respectively at early and very early epochs of cosmic evolution. In this sense, a double inflationary scenario naturally emerges.
 \end{abstract}

\maketitle

\section{Introduction}
\label{uno}
Inflationary paradigm  was
introduced  to address problems and shortcomings  related to the self-consistency of the Cosmological Standard Model at early stages of its evolution  \cite{Sta80,Guth81,Sato81p1,Sato81p2,Kazanas80}. The inflationary mechanism  can be achieved  in several different ways considering primordial scalar fields or geometric corrections into the effective gravitational action. The general aim is to address  problems like the  initial singularity,  the cosmological horizon, the cosmic microwave background isotropy (and the related anisotropies generated, in principle, with initial quantum fluctuations), the large scale structure formation and evolution, the absence of magnetic monopoles and so on \cite{Kolb,Mukhanov81,Guth82,Hawking82,Sta82}. The main ingredient of all these scenarios  is to claim that an inflationary phase occurs at some stage in the early
universe and that one  or more sources, different from standard ordinary matter,  give rise to accelerated cosmic expansion. Such an expansion can be a single or a multiple event often  related to the formation of structure at large and at very large scale.
In general, inflationary scenarios   originated from  some fundamental theory like  quantum gravity,  strings, M-theory or GUT models. Reversing the argument, inflationary models and observables related to inflation can be used to probe fundamental theories (see,  for example the latest results of the PLANCK and BICEP2 collaborations \cite{BICEP2, Planckdata, Planck2015}).

In particular, quantum fluctuations of
a given  scalar field  {\it i.e.} the   {\it inflaton},  gives  a mechanism for the origin of large scale structure. In other words,   inflation gives rise to  density perturbations that exhibit a scale invariant spectrum.  Such a feature, in principle,   is directly observed by measuring the temperature
anisotropies in cosmic microwave background \cite{Planckdata,WMAP,Peiris03,Tegmark03,Tegmarkdata,2dF}. 
A part the general features, the possibilities to realize inflation are several.
For example, in  the {\it old inflation},   inflaton is trapped in a false vacuum phase through a first order transition, while,  in the {\it new inflation},  expansion ends up with a  second order phase transition  after a slow rolling phase \cite{Guth81,Sato81p1,Sato81p2}. According to the problems to address, there are several different  inflationary models, for example the power law inflation, the hybrid inflation,  the oscillating inflation, the trace-anomaly driven inflation, the $k$-inflation, the ghost-inflation, the tachyon inflation and so on  \cite{Kolb99,Linde83,Linde82,Albre82,Freese90,Polarski92,Linde94}. Furthermore,  some of these models have no potential minimum and the inflationary mechanism appears different with respect to the standard one. See for example the quintessential inflation \cite{PV99} or the tachyon inflation \cite{FT02,Fe02,Paddy02,Sami02,SCQ02,TW05}. 

A natural way to achieve inflation is considering higher-order curvature corrections in the  Hilbert-Einstein Lagrangian \cite{PhysRepnostro,OdintsovPR,5,6,Mauro,faraoni, 9,10,libri,libroSV,libroSF}.  The first and well-known example of this approach is the Starobinsky model \cite{Sta80} where inflation is essentially driven by $R^2$ contributions, being $R$ the Ricci curvature scalar.  After this preliminary model, other  higher-order curvature terms have been taken into account  \cite{MO04,NOZ00,NO00,NO03,HHR01,ellis99,BB89,Maeda89}.
The philosophy is that, in the early higher-curvature regime, such further curvature invariants come out as renormalization terms in  quantum field theories in curved spacetime \cite{Birrell}. Furthermore, under conformal transformations, the theory  becomes minimally coupled in the Einstein frame. In this frame, the conformal scalar field assumes the role of  {\it inflaton} and leads the primordial  acceleration \cite{Maeda89}. However, more than one scalar field can be achieved by  conformal transformations disentangling the degrees of freedom present in the Jordan frame.  

Several combinations of curvature invariants, like $R_{\mu\nu}R^{\mu\nu},R_{\mu\nu\sigma\rho}R^{\mu\nu\sigma\rho}$ can be considered  
\cite{gorbu,ratbay,lorenzo1,lorenzo2}. The goal is   to explain both  the early and the late-time acceleration in a geometrical way \cite{capcurve} without invoking huge amount of dark energy or, sometime, ill-defined scalar fields. 
Among these attempt, a key role is played by the Gauss-Bonnet topological invariant  ${\cal G}$ that naturally arises in the process of quantum field theory regularization and renormalization  in  curved spacetime \cite{Birrell}. In particular, it contributes to the trace anomaly where higher-order curvature terms are present \cite{Barth}. In some sense, considering a theory where both $R$ and ${\cal  G}$ are nonlinearly present exhausts  the budget of  curvature degrees of freedom needed to extend General Relativity since the Ricci scalar and both the Ricci and the Riemann tensors are present in the definition of ${\cal G}$. From the inflation point of view, introducing ${\cal G}$ beside $R$ gives the opportunity to achieve a double inflationary scenario where the two acceleration  phases are led by 
${\cal G}$ and $R$ respectively. As we will see below, this happens as soon as both $R$ and ${\cal G}$ appear in non-linear combinations since linear $R$ means just General Relativity (and then no inflation) and linear ${\cal G}$ identically vanishes in  4D gravitational action, being an invariant. On the other hand, the combination of both terms seems to improve the inflationary mechanism since one achieves a $R$-dominated phase and a ${\cal G}$-dominated phase. The second  leads the Universe at very early stages of its evolution because ${\cal G}$ is quadratic in curvature invariants and then it  is dominant in stronger  curvature regimes. Specifically, using a non linear function of ${\cal G}$, inserted into the $f(R)$ approach, that is  
 a $F(R,{\cal G})$ function, {\it extends} the Starobinsky model since  the whole curvature "interactions",  present in the early Universe, are taken into account.  In view  of the recent results by the PLANCK  \cite{Planck2015} and BICEP2 \cite{BICEP2} collaborations, the potential advantages of this class of models, with respect to the original Starobinsky one, could be that curvature degrees of freedom (in particular the scalaron $R$) result better constrained (see \cite{Planckdata} for a detailed discussion). A first study in this sense is in the paper by Ivanov and Toporensky \cite{ivanov}, where cosmological dynamics  of fourth order gravity is studied in presence of Gauss-Bonnet term.

In this paper, we discuss the possibility to obtain inflation considering a generic $F(R,{\cal G})$ theory where, in principle, both $R$ and ${\cal G}$ are non linear in the action.  There are several recent studies on  models of this type \cite{OdiGB,defelice,defelice1,fGBnoether,antonio,topoquint,diego,myrz}. All of them, put in evidence the fact that the Gauss-Bonnet topological invariant can solve some shortcomings of the original $f(R)$ gravity and  contributes, in non trivial way, to the  accelerated expansion.

 The paper is organized as follows. In Sec. \ref{due},  we derive the field equations for $F(R, {\cal G})$.  General features of $F(R, {\cal G})$ cosmology and inflation are discussed in Sec. \ref{tre}. 
Sec \ref{quattro} is devoted to the discussion of  exact solutions coming from Noether symmetries giving rise to  power-law inflation.
Summary and  outlook are given in Sec. \ref{cinque}.

\section{Field equations of $F(R,\cal{G})$-gravity}\label{due}

Let us start by writing the most general action  for modified Gauss-Bonnet gravity
\begin{equation}
{\cal S}=\frac{1}{2\kappa}\int d^4x \sqrt{-g}F(R,{\cal G})\,,
   \label{action}
\end{equation}
where, as we said before,  $F(R,{\cal G})$ is a function of the Ricci scalar and Gauss-Bonnet invariant defined as
\begin{equation}
{\cal G}\equiv
R^2-4R_{\alpha\beta}R^{\alpha\beta}+R_{\alpha\beta\rho\sigma}
R^{\alpha\beta\rho\sigma}\,.
   \label{GBinvariant}
\end{equation}
Moreover $\kappa=8\pi G_N$, with $G_N$ Newton constant. We are using physical units $c= k_B=\hbar=1$.
We are discarding, for the moment,  the contribution of standard matter Lagrangian ${\cal L}_m$ that we will reconsider below.
The variation of the action (\ref{action}) with respect to the metric provides the following gravitational field equations \cite{antonio}
\begin{eqnarray}\label{eom}
&&G_{\mu \nu}= \frac{1}{ F_R} \Biggr[\nabla_\mu \nabla_\nu F_{R}-g_{\mu \nu} \Box F_{R}+2R \nabla_\mu \nabla_\nu F_{G}\nonumber\\&&
-2g_{\mu \nu} R \Box F_{G}-4R_\mu^{~\lambda} \nabla_\lambda \nabla_\nu F_{G}-4R_\nu^{~\lambda} \nabla_\lambda \nabla_\mu F_{G}  
\nonumber \\&&
+4R_{\mu \nu} \Box F_{G}+4 g_{\mu \nu} R^{\alpha \beta} \nabla_\alpha \nabla_\beta F_{G} 
+4R_{\mu \alpha \beta \nu} \nabla^\alpha \nabla^\beta F_{G}
 \nonumber \\&&
-\frac{1}{2}\,g_{\mu \nu}\bigr( R  F_R + {\cal G}  F_{\cal G}- F(R,{\cal G})\bigr) \Biggr]\,.
\end{eqnarray}
The trace is
\begin{equation} \label{trace}
3\left[\Box F_R + V_R\right]+R \left[\Box F_{\cal G}+ W_{\cal G}\right]=0,
\end{equation}
where  $\Box$ is the d'Alembert operator in curved spacetime and 
\begin{equation}
  \label{eq:def1}
  F_{R}\equiv\frac{\partial F(R,{\cal G})}{\partial R}\,,\qquad F_{\cal G}\equiv \frac{\partial F(R,{\cal G})}{\partial {\cal G}}\,, 
\end{equation}
are the partial derivatives with respect to $R$ and $\cal G$. 
It is possible to define two different potentials that depend on the scalar curvature  and the Gauss-Bonnet invariant that enter the trace equation with their partial derivatives
\begin{eqnarray}\label{pot}
&&V_R=\frac{\partial V}{\partial R}=\frac{1}{3}\left[R F_R-2 F(R, \cal G) \right], \\
\nonumber\\
&& W_{\cal G}=\frac{\partial W}{\partial {\cal G}}=2 \frac{{\cal G}}{R} F_{\cal G}.\end{eqnarray}
It is important to emphasize that, from  Eqs.(\ref{eom})-(\ref{trace}),  General Relativity is recovered as soon as $F(R,{\cal G})=R$. Furthermore, if ${\cal G}$ is not considered, we are exactly in the  $f(R)$ gravity context. Clearly, as  in the case of the Starobinsky $R$ {\it scalaron}, ${\cal G}$ plays the role of a further scalar field  whose dynamics is given by the Klein-Gordon-like Eq. (\ref{trace}). This means that we can expect a natural double inflation where both geometric fields play a role. As for the $R$ scalaron, we can expect a {\it mass} for the ${\cal G}$ scalaron which determine the "strength" of  the ${\cal G}$-dominated inflation.

\section{$F(R,{\cal G})$ double inflation}
\label{tre}

Let us consider now a  flat Friedman-Robertson-Walker (FRW)  metric
\begin{equation}
ds^{2}=-dt^{2}+a^{2}(t)(d{x}^{2}+d{y}^{2}+d{z}^{2})\,,
\label{metric}
\end{equation}
where $a(t)$ is the scale factor of the Universe. Inserting this metric into the action (\ref{action})  and assuming suitable Lagrange multipliers for $R$  and ${\cal G}$, 
we obtain the  point-like Lagrangian \cite{fGBnoether}
\begin{widetext}
\begin{eqnarray}
 {\cal L}&=& 6 a {\dot a}^2 F_R + 6 a^2  {\dot a}  {\dot F}_R-8  {\dot a}^3  {\dot  F}_{\cal G}
 +a^3\left[ F(R,{\cal G})-R\, F_R -{\cal G}  F_{\cal G}\right]\,,\label{PointLagra}
\end{eqnarray}
\end{widetext}
which is a canonical function depending on $t$ and  defined in the configuration space  ${\cal Q} \equiv\{a, R, {\cal G}\}$.
Specifically, the Lagrangian (\ref{PointLagra}) has a canonical form thanks to the Lagrange multipliers 
\begin{eqnarray}
R = 6 \left(2H^{2}+\dot H \right),
\label{eq:R}\end{eqnarray}
 \begin{eqnarray}
{\cal G} = 24H^{2} \left( H^{2}+\dot H \right), \label{eq:G}
\end{eqnarray}
that are also field equations for the related dynamical system \cite{fGBnoether}.
Here $\displaystyle{H=\frac{\dot a}{a}}$ is the Hubble parameter and the overdot denotes the derivative with respect to the cosmic  time  $t$. 
The cosmological  equations in term of $H$, are 
\begin{eqnarray}\label{FRWH}
&&\dot{H}=
\frac{1}{2 F_R + 8H \dot{F}_{\cal G}} \left[H \dot{F}_R-\ddot{F}_R+4H^3\dot{F}_{\cal G}-4H^2\ddot{F}_{\cal G}\right],\nonumber\\
\\
\label{energy}
&&H^2=
\frac{1}{6\;F_R+24H\dot{F}_{\cal G}}\left[F_RR-F(R,{\cal G})-6H\dot{F}_R+{\cal G}F_{\cal G}\right]\,,\nonumber\\
\end{eqnarray}
where Eq. \eqref{energy} is the energy condition, that is the $(0,0)$ Einstein equation.
The full dynamical system of $F(R,{\cal G})$ cosmology is given by Eqs. (\ref{eq:R}), (\ref{eq:G}), (\ref{FRWH}), (\ref{energy}).

To obtain  inflation,   the following conditions have to be satisfied: 
\begin{equation}
\left |\frac{\dot{H}}{H^2}\right|\ll1\,,  \hspace{1cm} \left |\frac{\ddot{H}}{H \; \dot{H}}\right|\ll1\,.
\end{equation}
It means, that the magnitude of the \textit{slow-roll parameters}
\begin{equation}
\epsilon=-\frac{\dot{H}}{H^2}, \;\;\;\;\;\;\; \eta=-\frac{\ddot{H}}{2\;H\;\dot{H}}\,,
\end{equation}
has to be small during inflation. Moreover, $\epsilon >0$ is necessary  to have $H<0$. The acceleration
is expressed as
\begin{equation}
\frac{\ddot{a}}{a}= \dot{H}+H^2\,,
\end{equation}
and then  the accelerated expansion ends only when the slow-roll parameter $\epsilon$ is of the unit order.

In order to discuss a possible  inflationary scenario, let us choose, for example, the following Lagrangian
\begin{equation}\label{starotopo}
{F(R,{\cal G})}= R+\alpha R^2+\beta {\cal G}^2\,,
\end{equation}
where $\alpha$ and $\beta$ are constants of the dimension length squared and length to the fourth
power respectively. 
The linear term in  $R$ is included to produce the correct weak-field limit.
It is easy to see that we have considered a $R^2$ model with a correction which adds new degrees of freedom due to the presence of the  Gauss-Bonnet term. 
In  the above Lagrangian, the term ${\cal G}^2$ is the first significant term  in ${\cal G}$ since the linear one gives no contribution\footnote{In four dimensions, we have\begin{equation}\label{note} \int d^4x\sqrt{-g}\,{\cal G}=0.\end{equation} This means that only a function of the Gauss-Bonnet invariant makes this integral non-trivial.
On the other hand, in five or higher dimensions Eq.\eqref{note} is different from zero. }.
 As it is well known, a theory like $f(R)=R+\alpha R^2$ is capable of producing an inflationary scenario \cite{Sta80} not excluded from the last PLANCK release \cite{Planck2015}.
 
Here we concentrate on the question if such an  inflationary scenario  can be improved considering the whole curvature budget that can be encompassed by adding a non linear function of the Gauss-Bonnet invariant.  In such a case, as stressed above, we can have a $R$-driven inflation led by the $R^2$ term and a ${\cal G}$-driven inflation led by ${\cal G}^2$ term. However, this is nothing else but a toy model that should be improved by realistic forms of the $F(R,{\cal G})$ function.

\begin{figure}[h!]
\thicklines
\includegraphics[scale=0.37]{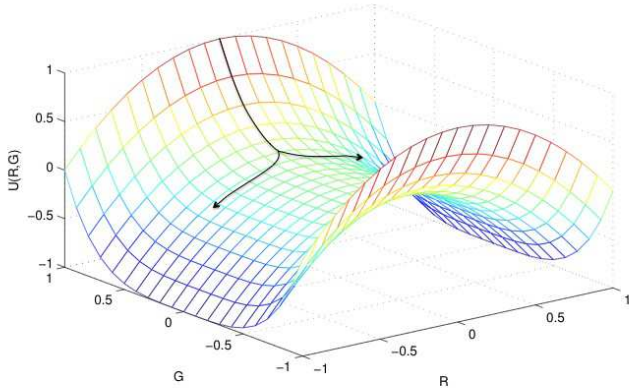}
\caption {Plot of $U(R,{\cal G})=\alpha R^2+\beta {\cal G}^2$. We note that the two fields can both  cooperate to the slow rolling phase. We assumed $\alpha$ and $\beta$ of the order unit with negative $\alpha$ and positive $\beta$. The choice of negative $\alpha$ is due to the stability conditions for the $R^2$ model discussed in \cite{barrow}. }
\label{fig1}
\end{figure}

To develop our considerations, let us consider the point-like Lagrangian (\ref{PointLagra}).
It is well known that, in analytical  mechanics,  any  Lagrangian can be decomposed as
\begin{eqnarray}
 L=K(q_i, {\dot q}_j)-U(q_i)\,,
\end{eqnarray}
where $K$ and $U$ are the kinetic energy and potential energy
respectively. Here we have   $q_i\equiv\{a,R,{\cal G}\}$ and    ${\dot q}_j\equiv\{{\dot a},{\dot R},{\dot {\cal G}}\}$.
In the case of Lagrangian  \eqref{PointLagra}, considering the Lagrangian density, i.e. ${\cal L}=a^3L$, it is 
\begin{eqnarray}
 K(a,{\dot a}, R, {\dot R}, {\cal G}, {\dot{\cal G}})= 6 \left(\frac{ {\dot a}}{a}\right)^2 F_R + 6 \left(\frac{ {\dot a}}{a}\right) {\dot F}_R-8 \left(\frac{ {\dot a}}{a}\right)^3  {\dot  F}_{\cal G}\,,\nonumber\\
 \end{eqnarray}
\begin{eqnarray}
 U(R,{\cal G})=-\left[ F(R,{\cal G})-R\, F_R -{\cal G}  F_{\cal G}\right]\,.
 \label{potenerg}
\end{eqnarray}
Assuming  the model \eqref{starotopo}, we have

\begin{eqnarray}
  L&=& \overbrace{6 \left(\frac{ {\dot a}}{a}\right)^2 \left(2 \alpha  R+1\right)+12 \alpha  \left(\frac{ {\dot a}}{a}\right){\dot R}-16 \beta \left(\frac{ {\dot a}}{a}\right)^3 {\dot{\cal G}}}^{\mbox{kinetic energy}}
\nonumber\\ \nonumber\\&&
 -\underbrace{\left[ \beta  {\cal G}^2+\alpha  R^2\right]}_{\mbox{potential energy}}\,,\label{PointLagraf}
\end{eqnarray}
 In Fig. \ref{fig1}, a qualitative  shape of the potential $U(R,{\cal G})$  is reported. A possible slow-roll trajectory is shown.

\begin{figure}[h!]
\thicklines
\includegraphics[scale=0.50]{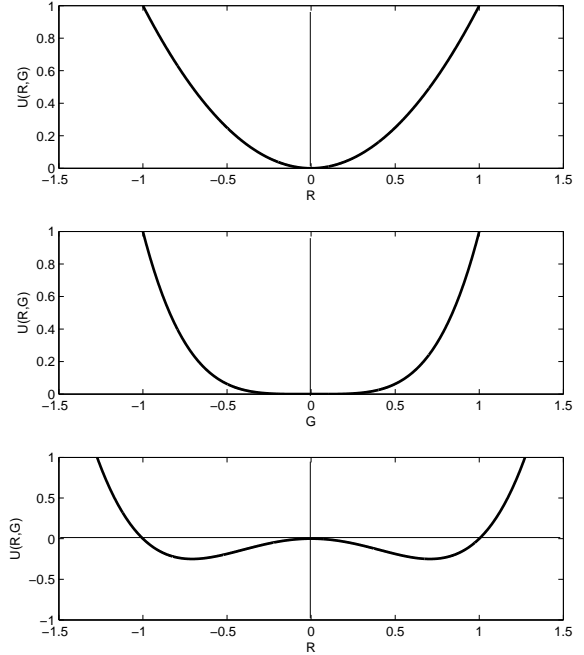}
\caption {Plots of  sections of the potential $U(R,{\cal G})=\alpha R^2+\beta {\cal G}^2$. In the top panel,  is reported the section of the potential when the $R^2$ term is dominant. In the central panel,   the case  where the term ${\cal G}^2$ is dominant. In bottom panel,  there is the behavior  of  $U(R,{\cal G})=\alpha R^2+\beta {\cal G}^2\sim \alpha R^2+\beta R^4$  whit respect to the Ricci scalar.  It is evident a  symmetry breaking and  a phase transition. \textbf{The values  of  $\alpha$ and $\beta$  are the same as in Fig.1}.}\label{fig2}
\end{figure}
It is important to stress the effective  behavior of the Lagrangian \eqref{starotopo} that assumes the following form 
\begin{equation}
F(R) \simeq R + \alpha R^2 + \beta R^4\,.
\end{equation}
In other words,  the correction to the $R^2$ model due to the presence of topological ${\cal G}^2$ term can be seen as  a sort of $\sim R^4$ correction. However, it is important to stress that ${\cal G}^2$ and $R^4$ have roughly the same dynamical role only at background level for the homogeneous and isotropic FRW metric. As soon as one takes into account anisotropies and inhomogeneities,  ${\cal G}^2$ and $R^4$ assume different roles since extra diagonal components of the Ricci and Riemann tensors cannot be discarded. In other words, considering the definition of the Gauss-Bonnet invariant, given in   Eq.(\ref {GBinvariant}), ${\cal G}\sim R^2$ only in the  FRW context. If more general situations are assumed, this approximation no longer holds. This means that ${\cal G}^2$ and $R^4$ can be observationally distinguished only evaluating anisotropies and inhomogeneities resulting from perturbations where extra diagonal components of the Ricci and Riemann tensors are not negligible.

In Fig. \ref{fig2}, the trends of $U(R,{\cal G})$ sections are reported according to the dominance of the  terms in the potential.

Let us describe now  the qualitative  evolution of the model. The behavior is  different depending on the "strength" of 
$R^2$ or ${\cal G}^2$ terms. In fact,  they give rise to a potential  with  two minima  that can be separated by a barrier  (see Fig. 2 in the bottom).  This represent a double inflationary scenario  where  ${\cal G}$-scalar dominates at early epochs,  at moderate early epochs dominate $R$-scalar and finally the model converges towards standard General Relativity. Due to the fact that ${\cal G}$ runs as ${\cal G}\simeq R^2$,  it is dominant at very high curvature improving, in some sense,  the Starobinsky inflation. In  the present simple toy-model,  we considered ${\cal G}^2$ and this means, as pointed out above,  that ${\cal G}^2\sim R^4$. From the energy condition, given by Eq. 
\eqref{energy}, we have
\begin{eqnarray}\label{Hquad}
&&12 \alpha H \ddot{H}+ H^2+36 \alpha H^2 \dot{H}+ 288 \beta H^4 \dot{H}^2 \\ \nonumber
&&+192 \beta H^5 \ddot{H} + 576 \beta H^6 \dot{H}-96 \beta H^8 - 6 \alpha \dot{H}^2=0\,,
\end{eqnarray}
and  from  \eqref{FRWH}, we obtain

\begin{eqnarray}\label{Hdot}
&& 576\beta H^2{\dot H}^3+768\beta H^3{\dot H}{\ddot H}+\beta H^4\left(1728{\dot H}^2+96{\dddot H}\right)
\nonumber\\&&+288\beta H^5{\ddot H}-384\beta H^6 {\dot H}^2
\nonumber\\&&+18\alpha H {\ddot H}+24 \alpha {\dot H}^2+6\alpha {\dddot H}+{\dot H}
=0\,.\nonumber\\
\end{eqnarray}
Considering the slow-roll conditions ${\dot H}<<H^2$ and ${\ddot H}<<H{\dot H}$,  this implies that $\displaystyle{\frac{\ddot H}{H}<<{\dot H}}$. From Eq.\eqref{Hquad}, one has
\begin{eqnarray} \label{Hquad1}
&& H^2+6\alpha\left(2H{\ddot H}+6H^2{\dot H}-{\dot H}^2\right)\nonumber\\&&
+96\beta H^4\left(3{\dot H}^2+2H{\ddot H}+6H^2{\dot H}-H^4\right)=0\,.
\end{eqnarray}
In order to study the evolution of the model, we have to distinguish among the various regimes.  Let us  suppose  that 
\begin{eqnarray} 
6\alpha>> 96\beta H^4
\end{eqnarray}
Then Eq. \eqref{Hquad1} takes the form
\begin{eqnarray} 
H^2+6\alpha\left(2H{\ddot H}+6H^2{\dot H}-{\dot H}^2\right)\cong0
\end{eqnarray}
and we obtain that 
\begin{eqnarray} 
m^2_R=\frac{1}{6\alpha}
\end{eqnarray}
and the solution for the scale factor is
\begin{equation}
a(t) \sim \exp \left[{\frac{t}{ \sqrt{6 \alpha}}}\right].
\end{equation}
This is nothing else but the well known $R^2$ inflation where the sign and the value of $\alpha$ determine the number of e-foldings \cite{vilenkin}.

On the other hand we can consider the regime   
\begin{eqnarray} 
96\beta H^4 >>   6\alpha\,,
\end{eqnarray}
where
\begin{eqnarray} 
H^2+96\beta H^4\left(3{\dot H}^2+2H{\ddot H}+6H^2{\dot H}-H^4\right)\cong0\,.
\end{eqnarray}
Inflation is recovered for 
\begin{eqnarray} 
 H^6\sim\frac{1}{96\beta}\,,
\end{eqnarray}
and then it is 
\begin{equation}
a(t) \sim \exp\left[{\frac{t}{ \sqrt[6]{96 \beta}}}\right]\,.
\end{equation}
From the above considerations,  we can  introduce a further mass term 
\begin{eqnarray} 
m^2_{\cal G}=\frac{1}{2\sqrt[3]{12 \beta}}\,,
\end{eqnarray}
due to the Gauss-Bonnet correction that leads another earlier inflationary behavior.
In conclusion, it seems that considering the whole curvature budget in the effective action ({\it i.e.} the further  combinations of curvature invariants more than the linear $R$) means to introduce two effective masses that lead the dynamics. It is important to stress that the parameters $\alpha$ and $\beta$ have to be consistent with the Solar System constraints according to the chameleon mechanism. Clearly, in the low energy regime,  General Relativity has to be recovered and then the quadratic and quartic terms in $R$ must be negligible. Essentially, starting from very early epochs, one has first to recover the Starobinsky model and then the Einstein regime. This means that  the two-scalaron regimes,  leading the two early inflationary phases, have to become negligible for $R\rightarrow 0$  to recover the standard Newtonian potential. In such a case, an  analysis like that in \cite{tsujikawa, nojiri} leads to assume the values of the parameters $\alpha$ and $\beta$ of the order unit  to achieve  the consistency  with the chameleon mechanism and   the Solar System experiments.

\section{$F(R,{\cal G})$ power-law Inflation}
\label{quattro}
Also power-law inflation can be easily achieved in the framework of $F(R,{\cal G})$ gravity.
In particular,  using the Noether Symmetry Approach  \cite{cimall} in the generic action (\ref{action}) and choosing appropriate Lagrangian multipliers that make the point-like Lagrangian canonical, models where conserved quantities emerge can be selected (see also \cite{safe,porco,35,36,fRTnoether,BDInoether} for analogue cases). 
This  means to impose 
\begin{eqnarray} 
L_X {\cal L}\,=\,0 \qquad \rightarrow \qquad X{\cal L}\,=\,0\,,
\label{LX}
\end{eqnarray}
where $L_X$ is the Lie derivative with respect to the Noether vector $X$ acting on the point-like Lagrangian ${\cal L}$.
A  possible choice  is to consider the  class of Lagrangians 
\begin{equation}\label{case1}
F({R,\cal G})= F_0 R^{n} G^{1-n},
\end{equation}
related to the presence of the Noether symmetries \cite{fGBnoether}.
For $n=2$, it is   $F({R,\cal G})= F_0 R^{2} G^{-1}.$ Inserting this choice into
 the point-like Lagrangian (\ref{PointLagra}), it  becomes,
\begin{eqnarray} 
{\cal L}&=&\frac{4\, F_0\, {\dot a}}{\cal G}\left[3\, a\, {\dot a}\, R+3\,a\,{\dot R}-3\, a^2 \,{\dot {\cal G}} \left(\frac{R}{\cal G}\right)\right.\nonumber\\&&\left.+4\, {\dot a}^2 \,{\dot R}\left(\frac{R}{\cal G}\right)-4\, {\dot a}^2\,{\dot {\cal G}}\left(\frac{R}{\cal G}\right)^2\right]\,.
\label{lagrangian_n2}
\end{eqnarray}
The same choice can be done into the cosmological Eqs. (\ref{FRWH}) and (\ref{energy}) that are nothing else but the Euler-Lagrange equations of the Lagrangian (\ref{lagrangian_n2}) together with the Lagrange multipliers (\ref{eq:R}) and (\ref{eq:G}).
Power law solutions for \eqref{lagrangian_n2} are easily  found \cite{fGBnoether,topoquint}. 
For example we have
\begin{equation}\label{sol1}
 a(t)=t^s \, , \hspace{1cm} {\text{with}}  \;\;\;\;  n=2  \;\;\;\; {\text{and}}  \;\;\;\;\, s=3 \;.
\end{equation}
A further interesting solution is 
\begin{equation}\label{sol2}
a(t)=t^s \, , \hspace{1cm} {\text{with}}  \;\;\;\;  n=\frac{3}{4}  \;\;\;\; {\text{and}}  \;\;\;\;\, s=\frac{1}{2} \;.
\end{equation}
General conditions between the exponents $n$ and $s$ are:
\begin{equation}\label{conditions}
  n=\frac{1+s}{2}  \hspace{0,3cm}  {\text{and}} \;\; n=\frac{1}{1+2s(s-1)}-2s\;.
\end{equation} 
In one these conditions the two constraint are satisfied and are of the same form.
It is easy to verify that solutions \eqref{sol1} and \eqref{sol2} are in one of these cases.

\begin{figure}[h!]
\includegraphics[scale=0.50]{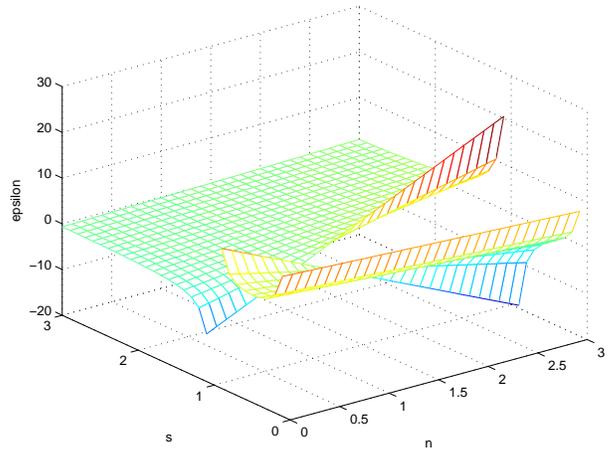}
\caption {Plot of $\epsilon(n,s)$. The allowed region for inflation is the green one, in that region the value of $\epsilon$ is less than 1.}\label{figep}
\end{figure}

\begin{figure}[h!]
\includegraphics[scale=0.50]{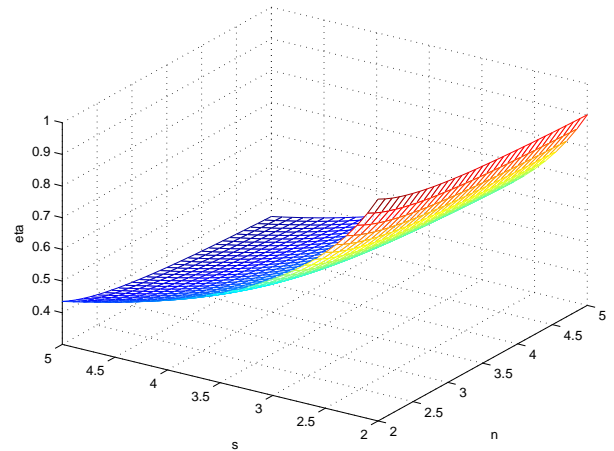}
\caption {Plot of $\eta(n,s)$ parameters. The blue and green part in the figure are the allowed solutions.}\label{figep1}
\end{figure}

In order to discuss inflation, we have to consider Eqs. \eqref{FRWH} and \eqref{energy}.  One obtains the following relations 
\begin{equation}
\dot{H}=-\frac{s(n-1) \left[ n (6s-4)-3s(s+1)+4 \right] }{\left[ s(s-5)+2n(2s-1)+2\right] t^2},
\end{equation}
\begin{equation}
H^2=-\frac{2s^2(s-1)(n-1) }{\left[ s(s-5)+2n(2s-1)+2\right] t^2}\,.
\end{equation}
A condition for inflation is 
\begin{equation}
\left|\frac{\dot{H}}{H^2}\right|=\left|\frac{2s(n-1)-2(s-1)+s(s-1)}{2s(s-1)}\right| \ll 1\,.
\end{equation}
The slow roll conditions are 
\begin{equation}
   \epsilon=\frac{2s(1-n)+2(s-1)-s(s-1)}{2s(s-1)} \ll 1\,,
\end{equation}
\begin{equation}
   \eta =\frac{1}{\sqrt{2}}\sqrt{\frac{s^2(s-1)(n-1)}{s(s-5)+n(4n-2)+2}}  \ll 1\,.
\end{equation}
Considering the relation ${\displaystyle n = \frac{(1 + s)}{2}}$, slow roll conditions on $\epsilon$ and $\eta$ are satisfied for $s> 2.171$.
In conclusion, we can easily see that, for relatively
large $s$,  slow-roll conditions are  satisfied. In Figs. \ref{figep} and \ref{figep1},  qualitative pictures of  the parameter space regions where inflation is allowed are reported.

Furthermore we can evaluate  the  anisotropies and the power spectrum coming from
inflation using the slow-roll parameters. The spectral index $n_s$ and the tensor-to-scalar ratio $r$ are respectively 
\begin{eqnarray}
n_s=1-6\epsilon+2\eta\,, \qquad r=16\epsilon\,,
\end{eqnarray}
 while the amplitude of the primordial power spectrum is
 \begin{eqnarray}
 \Delta^2_{\cal R}=\frac{\kappa^2H^2}{8\pi^2 \epsilon}\,.
\end{eqnarray}
We obtain that the values $n_s\sim 1.01$ and $r\sim 0.10$  are  in good agreement  with the observational values of spectral index estimated by  PLANCK$+$WP data, {\it i.e.} $n_s=0.9603\pm0.0073$ (68$\%$ CL) and $r<0.11$ (95$\%$ CL) \cite{Planckdata,Planck2015}. These results are  consistent also with the values measured by the BICEP2 collaboration 
\cite{BICEP2}. 

Finally, it is possible to estimate  the grow factor for  the class of models  $F({R,\cal G})= F_0 R^{n} G^{1-n}$. Let us consider the  equation which governs the evolution
of the matter fluctuations in the linear regime
\begin{equation}
\label{odedelta}
\ddot{\delta}_{m}+ 2H\dot{\delta}_{m}-4 \pi G_{\rm eff} \rho_{m} \delta_{m}=0\,,
\end{equation}
where $\rho_{m}$ is the matter density and
$G_{\rm eff}$ is the effective Newton coupling which, in our case,  is
\begin{eqnarray}
\label{Geff}
G_{\rm eff}=\frac{G_N}{F_{R}(R,{\cal G})} \,,
\end{eqnarray}
where  $G_N$ is the Newton gravitational constant. However, we are considering perfect fluid matter that enters minimally coupled in action \eqref{action}.
We use  Eq. \eqref{energy} with matter density contribution as follow
\begin{eqnarray}
\label{reff}
4\pi G \rho_{(m)}=\frac{3H^{2}}{2}-4\pi G \rho_{({\cal GB})}\,,
\end{eqnarray}
with
\begin{eqnarray}
\rho_{({\cal GB})}=\frac{R F_R-F(R,{\cal G})-6H\dot{F}_R+{\cal G}F_{\cal G}-24H^3\dot{F}_{\cal G}}{16\pi G_N}\,.\nonumber\\
\end{eqnarray}
Inserting Eqs. (\ref{Geff}) and (\ref{reff})  
into Eq.(\ref{odedelta}),  we obtain the  equation 
\begin{eqnarray}
\label{odedelta1}
&&\ddot{\delta}_{m}+ 2H\dot{\delta}_{m}+\nonumber\\&&+\frac{R F_R-F(R,{\cal G})-6H\dot{F}_R+{\cal G}F_{\cal G}-24H^3\dot{F}_{\cal G}}{4 F_R}\delta_{m}=0\,.\nonumber\\
\end{eqnarray}
Now, considering relations \eqref{conditions}, we have $a(t)=a_0t^s=a_0t^{2n-1}$ and consequently $\displaystyle{H=\frac{2n-1}{t}}$, therefore, Eq.(\ref{odedelta1}) becomes
\begin{eqnarray}\label{odedelta2}
{\ddot \delta}_m +\frac{2n-1}{t} {\dot \delta}_m+\frac{3(6n^2-6n-1)}{2t^2}\delta_m=0\,.
\end{eqnarray}
 Eq.(\ref{odedelta2}) is 
an Euler equation whose general solution is
\begin{equation}
\delta_{m} (t)= t^{\frac{1}{2} \left(-\sqrt{3-8 n^2}-4
   n+3\right)} \left(c_2 t^{\sqrt{3-8 n^2}}+c_1\right)\,.
\end{equation} 

\begin{figure}[h!]
\includegraphics[scale=0.45]{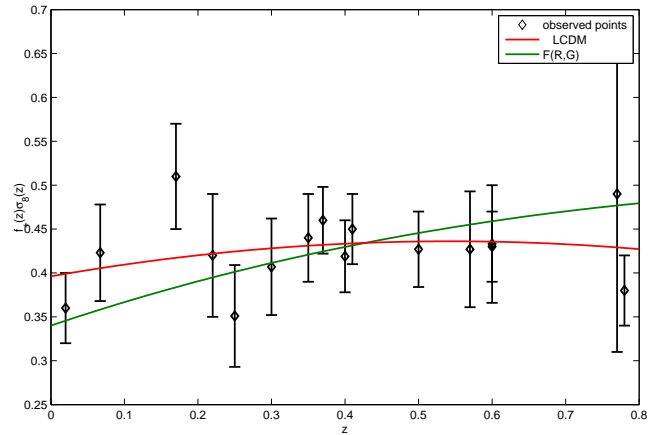}
\caption {The plot shows the comparison of the growth
rate $f_{+}(z)\sigma_{8}(z)$ for $F({R,\cal G})= F_0 R^{n} {\cal G}^{1-n}$ (green line) compared to that of $\Lambda$CDM (red line). The solid points are the observed one \cite{BasNes13}. For $F({R,\cal G})$ we consider  the value $n=2$. The parameter $F_0$ is assumed as a "prior" normalized at the $\Lambda$CDM value of the gravitational constant. This means that, in our units, it can be assumed of order unit. See also \cite{basilakos_noi}  for details.} \label{growfinal}
\end{figure}

Since $a(z)=(1+z)^{-1}$ we have that 
\begin{equation}
\label{HE}
H=H_{0}a^{-\frac{1}{2n-1}}=H_{0}\left(\frac{1}{1+z}\right)^{\frac{1}{2n-1}}\,,
\end{equation}
where $H_{0}$ is the Hubble constant that can be chosen as a prior in agreement with data.
The deceleration parameter $q$ is
\begin{equation}\label{eq:qnu}
q=-1-\frac{{\rm d\ln}H}{{\rm d \ln}a}=-1+\frac{1}{2n-1}\;.
\end{equation}
In Fig. \ref{growfinal}, the comparison between a  $F(R,{\cal G})$ model  with the $\Lambda$CDM analogue is reported.

By a rapid inspection of the figure, it is evident that there is  no change in the evolution of the curve  
since, for any  $F(R,{\cal G})=F_0 R^{n} {\cal G}^{1-n}$  model
 the deceleration parameter preserves sign, and therefore the
universe always accelerates or always decelerates
depending on the value of $n$. Clearly, for  $n=1$,
the  solution is an  Einstein-de Sitter model as it has to be. On the other hand, the 
accelerated expansion of the universe ($q<0$) is recovered for $n>1$, but, in this case, the universe accelerates forever without the possibility of structure formation.
In conclusion, we have to stress that more realistic models are necessary in order to fit the observations.

\section{ Conclusions}
\label{cinque}
In this paper, we have considered the possibility to obtain cosmological inflation starting from a generic function $F(R,{\cal G})$ of the Ricci curvature scalar $R$ and the Gauss-Bonnet topological invariant $\cal G$. Such a kind of theories, due to the algebraic relation among the curvature invariants in ${\cal G}$, see Eq. \eqref{GBinvariant},  can exhaust the whole curvature budget of effective gravitational theories where derivatives of curvature invariants are not present. The main feature that emerges by this approach is the fact that two effective masses have to be considered, one related to $R$ and the other related to ${\cal G}$. These masses define two different scales that drive dynamics at early and very early epochs, giving rise to a natural  double inflationary scenario. Here we have sketched the essential characteristics of this picture considering exponential and power law inflation.
However, the theory has to be  worked out in order to select reliable models to be compared with data. In a forthcoming paper,  the matching with data will be addressed in details.

\section*{Acknowledgements}
The Authors thank the Referee for the useful comments and hints that allowed to improve the paper.
The authors acknowledge INFN Sez. di Napoli (Iniziative Specifiche QGSKY, QNP, and TEONGRAV) for
financial support.


\end{document}